# Multimodal assessment of the skin biomechanical properties during wound healing and hyaluronic acid response

**Arthur BOUFFANDEAU [a], David POISBLEAU [a], Mohand Ou Idir OUIDJA [b], Patrick KARASINSKI [c], Dulce PAPY [b], Charles-Henri FLOUZAT-LACHANIETTE [d,e], Romain BOSC [d,f], Barbara HERSANT [d,f], Jean-Paul MENINGAUD [d,f] and Guillaume HAIAT [a]**

**[a] CNRS, Univ Paris Est Créteil, Univ Gustave Eiffel, UMR 8208, MSME, 61 Avenue du Général de Gaulle, 94010 Créteil, France.**

**[b] Univ Paris Est Créteil, Glycobiology, Cell Growth and Tissue Repair Research Unit (Gly-CRRET), 61 Avenue du Général de Gaulle, 94010 Créteil, France.**

**[c] Univ Paris Est Créteil, Univ Gustave Eiffel, CNRS, UMR 8208, MSME, 61 Avenue du Général de Gaulle, 94010 Créteil, France.**

**[d] INSERM U955, IMRB Université Paris-Est, 8 rue du Général Sarrail, 94010 Créteil, France.**

**[e] Service de Chirurgie Orthopédique et Traumatologique, Hôpital Henri Mondor AP-HP, CHU Paris 12, Université Paris-Est, 51 avenue du Maréchal de Lattre de Tassigny, 94000 Créteil, France.**

**[f] Plastic, Reconstructive, Aesthetic and Maxillofacial Surgery Department, Hôpital Henri Mondor AP-HP, CHU Paris 12, 51 avenue du Maréchal de Lattre de Tassigny, 94000 Créteil, France.**

**Corresponding author:**

Guillaume HAÏAT

Laboratoire de Modélisation et de Simulation Multi-Échelle, UMR CNRS 8208,
61 avenue du Général de Gaulle, 94010 Créteil, France.

tel: (33) 1 45 17 14 41
fax: (33) 1 45 17 14 33
e-mail: guillaume.haiat@cnrs.fr

## Abstract

Various methods such as elastography or mechanical testing have been developed to quantify the mechanical properties of soft tissues, which are related to the tissue local pathophysiological conditions. This study is the first one to investigate the *in vivo* performances of the Impact-Based Analysis Method (IBAM) to characterize the biomechanical behavior of skin during physiological mechanisms.

A murine model (12 Sprague-Dawley rats) was considered and various measurements of skin mechanical properties were conducted on different regions of interest over a 3-week period. The effects of wound healing after full-thickness excisional wound and an intradermal hyaluronic acid (HA) injection were monitored using IBAM, as well as two commercially available devices: Cutometer® and MyotonPro®.

The indicators $\Delta t$ (from IBAM), $R0$ and $R5$ (from Cutometer®) vary significantly as a function of wound healing time, which corresponds to a stiffening of healing tissue compared to healthy tissue. The indicator $\Delta t$ and $S$ (from MyotonPro®) are significantly different when HA was injected compared to healthy tissue, which also corresponds to less stiff tissue. Additionally, both IBAM and Cutometer® detect significant biomechanical variations in superficial soft tissue related to tissue growth in all measurement regions.

This first *in vivo* use of IBAM shows that IBAM could assess significant variations related to wound healing or regeneration mechanisms following HA injection. This study paves the way for the development of a simple, low-cost and objective device. Future clinical studies or other animal models should be carried out to confirm these results.

# 1. Introduction

The skin is the largest organ of the human body and is responsible for the protection of internal organs, the thermal regulation and the touch sensation. However, various endogenous factors (ethnicity, hormonal factors and anatomical location) and exogenous aggressions (lifestyle and UV exposure…) (Farage et al., 2008) may alter the properties of the different skin layers (epidermis, dermis and hypodermis). Therefore, it is essential to understand the physiological mechanisms at work in skin tissue, which will allow the development of new products and the improvement of skin care practices, both from clinical and aesthetic perspectives (Khavkin and Ellis, 2011). To do so, complementary to purely biological aspects, the skin has been studied from other perspectives, such as its biomechanical behavior (Singh and Chanda, 2021; Arnold et al., 2023), the thickness of the epidermis or dermis, its coloration, its moisture level or its transepidermal water loss (Nam et al., 2015; Nedelec et al., 2016; Charipoor et al., 2024).

The skin biomechanical behavior has emerged as an interesting source of information on its physiological state under different scenarios. For instance, changes in tissue elasticity have been observed during skin growth and aging (Cua et al., 1990; Ahn et al., 2007; Ryu et al., 2008; Kim et al., 2018). Mechanical properties are also good indicators to quantify differences between pathological and healthy cases. For example, they have been used to compare new treatments during the healing process (Held et al., 2016; Lintel et al., 2023), to observe photoaging effects relative to UV exposure (Moloney et al., 1992; Edwards et al., 2001; Imokawa and Ishida, 2015) and to monitor changes after intradermal injection of hyaluronic acid (Seok et al., 2016; Heo et al., 2021).

Although invasive, histology remains a reference method (Kanitakis, 2002; Khavkin and Ellis, 2011) from a clinical perspective to diagnose skin cancers. From the biomechanical point of view, microindentation and compression or tensile tests have been widely investigated (McKee et al., 2011; Arnold et al., 2023) in this regard. In the last two decades, various simple and non-invasive techniques have been developed to measure the biomechanical properties of soft tissues. These techniques include torsion (Serup et al., 2006) (e.g. Dermal Torque Meter® (Murray and Wickett, 1997)), suction (e.g. Cutometer® or Dermaflex® (Murray and Wickett, 1997; Serup et al., 2006)), indentation (Serup et al., 2006) (e.g. SkinFibrometer® (Kim et al., 2018)), vibration (e.g. Reviscometer® (Rosado et al., 2017)) and ballistic (e.g. Ballistometer® (Serup et al., 2006; Woo et al., 2014)) methods. In particular, Cutometer® is used to assess the viscoelastic properties of the epidermis or dermis of the skin (Serup et al., 2006). This suction technique has become the reference method in the cosmetics sector (Dobrev, 2005; Ryu et al., 2008). MyotonPro® is a hand-held device analyzing the tissue dynamic response to a mechanical stimulus (Bizzini and Mannion, 2003). It is widely used, in research only, to monitor the biomechanical behavior of superficial soft tissues (Feng et al., 2018; John et al., 2023).

More recently, our group has developed a new technique for the mechanical characterization of superficial soft tissues: the Impact-Based Analysis Method (IBAM) (Poudrel et al., 2024; Bouffandeau et al., 2025, 2026). This method derives from impact analysis techniques used to quantify i) the stability of cementless implants during hip arthroplasty surgery (Michel et al., 2015, 2016; Albini Lomami et al., 2020; Dubory et al., 2020) and ii) the osteotome progression in bone tissue during osteotomy procedures (Hubert et al., 2020; Lamassoure et al., 2021b, 2021a; Bas Dit Nugues et al., 2023; Bas dit Nugues et al., 2025). During an IBAM measurement, a punch in contact with the soft tissue is impacted by a hammer instrumented with a force sensor. The signal corresponding to the variation of the force as a function time during the impact is analyzed to determine the tissue elastic properties (Poudrel et al., 2024; Bouffandeau et al., 2025). The performances of IBAM were quantified and tested using soft tissue phantoms (Poudrel et al., 2024; Bouffandeau et al., 2025, 2026), including the repeatability (which corresponds to the variability under the same measurement conditions, i.e. without moving the set-up), the reproducibility (variability when the set-up is repositioned) and the sensitivity to variations of tissue stiffness. The stiffness sensitivity of IBAM was shown to be equivalent or better compared to that of Dynamic Mechanical Analysis (DMA) (Bouffandeau et al., 2025), a reference approach for materials characterization, and to that of elastographic techniques (Poudrel et al., 2024; Bouffandeau et al., 2025). Furthermore, in a comparative study involving Cutometer® and MyotonPro®, IBAM was shown to be the most effective technique for characterizing mechanical properties at relatively low depth, from 1 to 10 mm (Bouffandeau et al., 2026). Finally, it was proven that the IBAM can accurately predict certain mechanical properties using soft tissue mimicking phantoms (Poudrel et al., 2024; Bouffandeau et al., 2025, 2026). The force signal has been shown to depend on the Young's modulus of the characterized samples. IBAM enables an objective assessment of superficial soft tissue stiffness, as opposed to palpation. Additionally, the estimation error of IBAM was significantly better than those of elastographic techniques currently used in clinical practice, as well as those of the main simple and non-invasive techniques available on the market (Bouffandeau et al., 2026). The measurement process takes less than one minute. Moreover,

IBAM is a relatively simple device, allowing to consider the conception of a low-cost device. No special technical skills will be required to perform the measurement. However, IBAM has not yet been applied to biological tissues. Therefore, its relevance and viability must first be established through measurements conducted on animal tissues. These experiments will provide the necessary data to justify extending the validation of the method to human tissue. Murine models are commonly used as a first approximation of skin due to their simplicity and affordability, despite size differences in skin layers and the presence of an extra-muscular layer in mice, the panniculus carnosus (Cléry-Barraud et al., 2013; Zomer and Trentin, 2018).

The purpose of this paper is to determine the influence of wound healing and regeneration mechanisms (following hyaluronic acid injection) on the biomechanical behavior of the skin, based on various non-invasive mechanical surface characterization techniques. IBAM, Cutometer® and MyotonPro® were used to assess the elastic properties of skin tissue in a murine model over a 3-week period. For each animal, mechanical measurements were performed at skin sites exposed to wound healing and hyaluronic acid injection, and at control sites. Significant tendencies and differences of behavior were identified and compared for the different techniques.

# 2. Materials and Methods

## 2.1 Animals

The animal cohort consisted of 12 adult male Sprague-Dawley rats (Janvier Labs, Saint Berthevin, France). Restricting the cohort to males allowed to control the variability related to gender-based differences, such as skin morphology (Azzi et al., 2005). To minimize age-related differences, the animals were 6 weeks old at the beginning of the experiments. They were housed in groups of 2 in cages with unlimited access to food and water, and supplemented with cotton swabs. The cages were kept in a controlled environment in terms of temperature (between 21 and 23°C) and light cycle (12h light / 12h dark). The animals were acclimated for 1 week prior to the start of the protocol. Throughout the protocol, their conditions were monitored daily to detect pain and potential weight loss. The cutaneous conditions of the areas affected by full-thickness excisional wound and intradermal injection were also monitored daily. This animal experiment was conducted in accordance with the European Directive 2010/63/EU for care and use of laboratory animals and approved by the local ethical committee ComEth CE016 (n°#44667-2023061717481626 v7).

## 2.2 Protocol

As shown in Fig. 1, the mechanical properties of superficial soft tissues of five regions of interest were measured. The first couple of measurements monitored the effects of the wound healing process on the right upper back site, referred to as "wound healing". The results were compared with the symmetrical healthy region, in the left upper back site, referred to as "control". The second couple of measurements investigated the changes due to the tissue regeneration following hyaluronic acid injection on the right lower back site, referred to as "HA – 0.5mL" or "HA – 0.25mL". The results were also compared with the symmetrical control region, in the left lower back site, referred to as "control". The last set of observations corresponds to another control behavior to study the time evolution of the skin properties on the median ventral site, considered as healthy superficial soft tissue and away from previous sites involving physiological responses. For each experimental day (denoted $D_i$), the three characterization techniques (Cutometer®, MyotonPro® and IBAM) were employed to perform measurements for each of the 5 regions of interest and for each animal.

The first part of the experimental protocol, carried out on the first day of the protocol ($D_0$), consisted in characterizing the different sites under healthy conditions. Then, two lesions were performed for each rat: a full-thickness excisional wound and an intradermal injection of hyaluronic acid. Firstly, rats were anesthetized by inhalation of Isoflurane (Vetflurane®, Virbac, Carros, France) with a minimum alveolar concentration (MAC) of 3% for the induction phase and a MAC of 2% for the maintenance phase. This anesthesia was supplemented by an intraperitoneal injection of 90 mg/kg Ketamine (Virbac, Carros, France) + 10 mg/kg Xylazine (Rompun®, Bayer, Leverkusen, Germany) to minimize the risk of rats awakening during the experiment. Secondly, a shaving with clippers and an identification of the regions of interest with markers were performed before positioning the animal within the measurement support, as shown in Fig. 2. Then, initial mechanical measurements were performed in the five regions of interest (which were all still in healthy conditions) using the three characterization techniques. Thirdly, after careful disinfection with Dakin (Cooper, Melun, France), a full-thickness excisional wound to the subcutaneous layer of 10×10 mm was performed at the right upper back site using a scalpel for all rats. Meanwhile, an intradermal injection of hyaluronic acid (Hyamira® Forte, Nyuma Apharm, Usmate-Velate, Italy) of 0.50 mL for 6 rats and 0.25 mL for the other half of the group was performed at the right lower back site, referred to as "HA – 0.5mL" and "HA – 0.25mL", respectively. The anesthesia was then discontinued, and the wound site was treated with Vaseline (Cooper, Melun, France), and a bandage consisting of a sterile elastic bandage (Raffin, Saint Romain de Popey, France) and a surgical tape (3M Deutschland GmbH Health Care Business, Neuss, Germany) to protect the lesion, before placing the rat back in its cage.

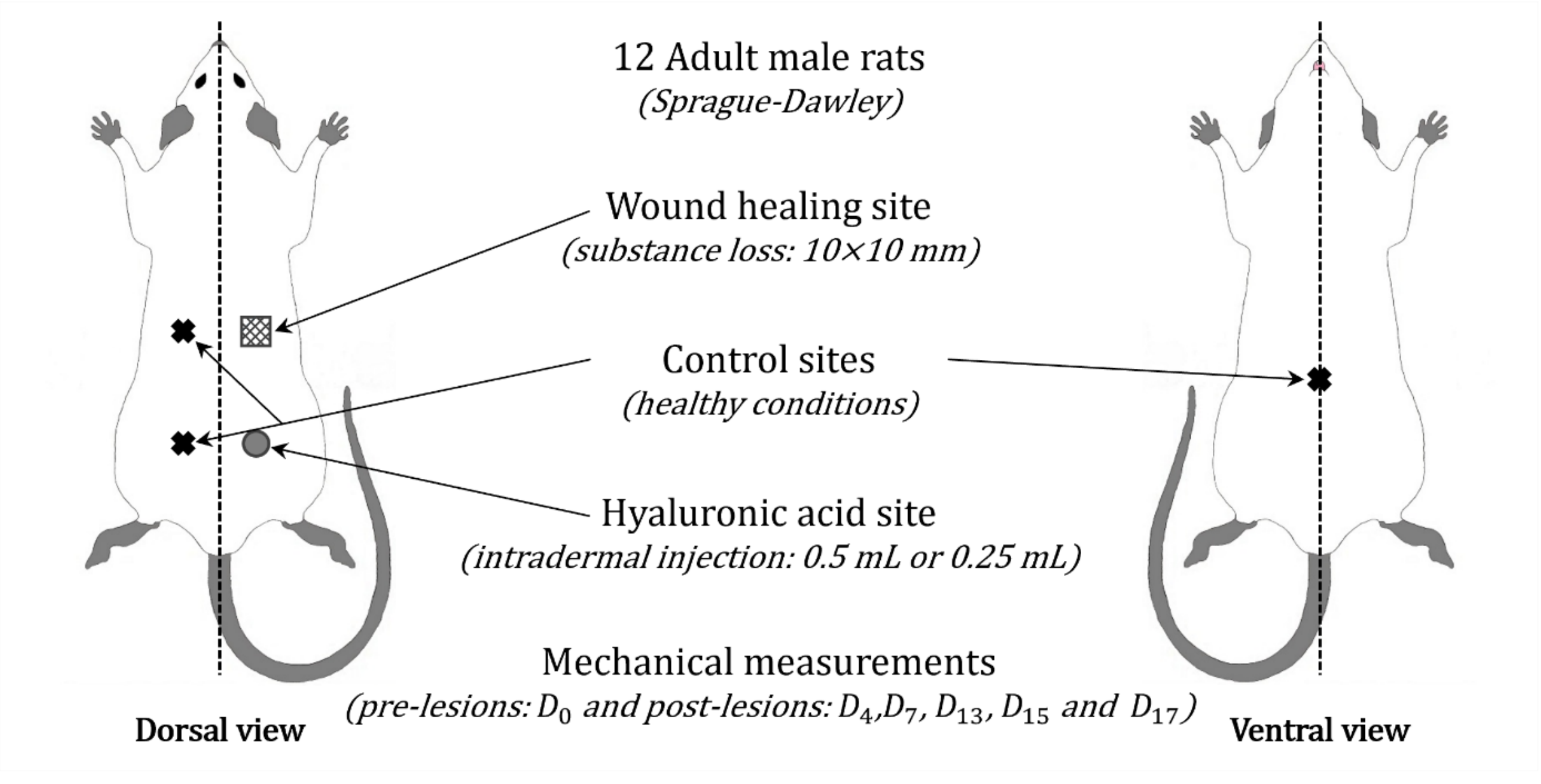

***Figure 1:*** *Schematic illustration of the experimental protocol. Location of the five sites ("wound healing", "hyaluronic acid injection" and 3 "control") characterized by the three mechanical surface characterization techniques on different experimental days, noted* $\boldsymbol{D_i}$.

The second part of the experimental protocol consisted in characterizing the mechanical properties of the different regions of interest as a function of time. Measurements were made using the three mechanical surface characterization techniques at 4, 7, 13, 15 and 17 days after the creation of the lesions, respectively denoted $D_4$, $D_7$, $D_{13}$, $D_{15}$ and $D_{17}$. Thus, for each $D_i$ ($i \in \{4, 7, 13, 15, 17\}$), the rats were anesthetized by inhalation of Isoflurane similar to the prior description. The different regions of interest were shaved with clippers if necessary and the wound healing region was disinfected with Dakin at $D_4$ and $D_7$. The rats were then positioned in the dedicated support, as shown in Fig. 2, in order to perform measurements at the five sites using the three characterization techniques. Finally, the anesthesia was stopped and before returning the rat to its cage, the wound healing region was again covered with Vaseline and a bandage at $D_4$ or with a hydrocolloid dressing (Comfeel® Plus Transparent, Laboratoires Coloplast, Fontenay-sous-Bois, France) between $D_7$ and $D_{15}$.

Supplementary repeatability and reproducibility values were estimated for the three mechanical surface characterization techniques at the end of the protocol. For each characterization technique, 5 successive measurements were taken without any displacement of the animal, the support or the instrumentation between measurements (corresponding to repeatability). This last series of measurements was then repeated 5 times after repositioning the animal and the device between each series. This reproducibility estimation was carried out on three distinct regions of interest: the left and right upper back sites, and the median ventral site. However, since the absolute values of repeatability and reproducibility depend on the absolute values of the indicators, these variability values are expressed as percentages, relative to the mean value of the indicators. Finally, these reproducibility values were used to calculate the confidence intervals of the mean difference described in subsection 2.4.

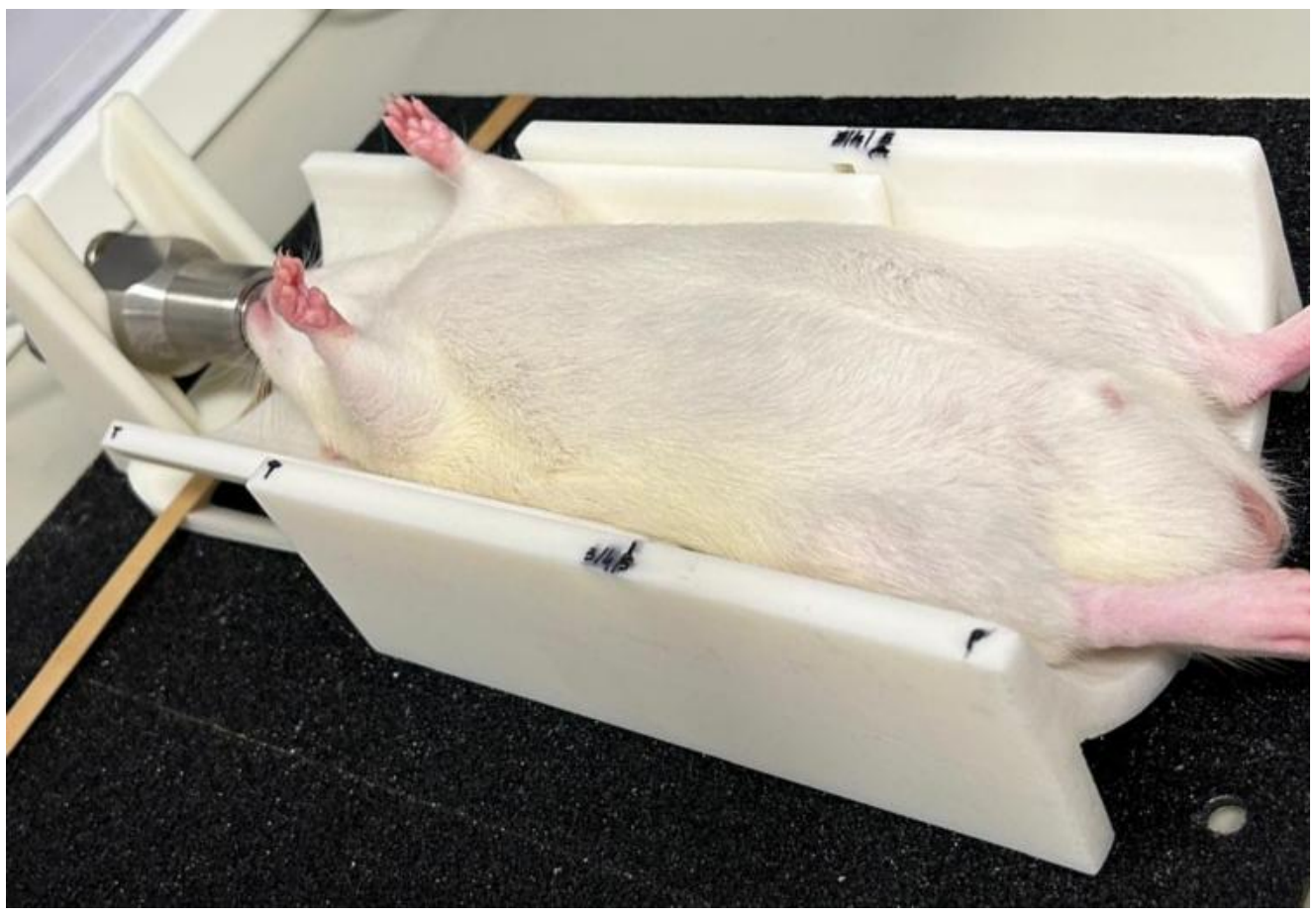

***Figure 2:*** *Picture of the positioning of an anesthetized rat in the measurement support before shaving and identification of measurement site at first day of the protocol* $\boldsymbol{D_0}$.

### 2.3 Mechanical surface characterization techniques

The soft tissues mechanical properties were investigated using Impact-Based Analysis Method (IBAM), as well as two other commercial characterization techniques: Cutometer® and MyotonPro®, following the results obtained *in vitro* (Bouffandeau et al., 2026). Prior to each measurement, the anesthetized rats were positioned within a support, as illustrated in Fig. 2, to ensure reproducibility between measurements as far as practicable. Several markers and a curved support shape were used to standardize the positioning of different animals relative to a reference posture.

**Impact-Based Analysis Method (IBAM)** aims at assessing the biomechanical properties of superficial soft tissues (Poudrel et al., 2024; Bouffandeau et al., 2025, 2026). During each IBAM measurement, a 3-gram hammer instrumented with a force sensor (type 8204, Brüel and Kjær (B&K), Naerum, Denmark) impacted a vertically guided punch in contact with the rat skin, thereby generating a force signal, as illustrated in Fig. 3A. The cylinder (referred to as the "punch" in what follows) is a 4 mm-diameter aluminum cylinder with a 2 mm-diameter hemisphere at its end in contact with the rat tissue. The force signal resulting from the impact #*i* was recorded using LabVIEW software (National Instruments, Austin, TX, USA) with a sampling frequency of 102.4 kHz (NI 9232, National Instruments, Austin, TX, USA). The times of the first two peaks $t_{i,1}$ and $t_{i,2}$ and the amplitude $F_{i,1}$ of the first peak were then determined. As shown in Fig. 3B, the first peak corresponded to the initial impact of the hammer on the punch, while subsequent peaks corresponded to the rebounds of the punch between the hammer and the soft tissue surface (Poudrel et al., 2024; Bouffandeau et al., 2025). Finally, for the impacts having $F_{i,1}$ values comprised between 8 and 10 N, an indicator $\Delta t_i = t_{i,2} - t_{i,1}$ was calculated and the final indicator $\Delta t$ was computed as the average of five consecutive impacts (corresponding to the average of five $\Delta t_i$ values) (Bouffandeau et al., 2025, 2026). Under this protocol, a measurement was composed of 5 impacts to provide a $\Delta t$ value for each region of interest, each rat and each day.

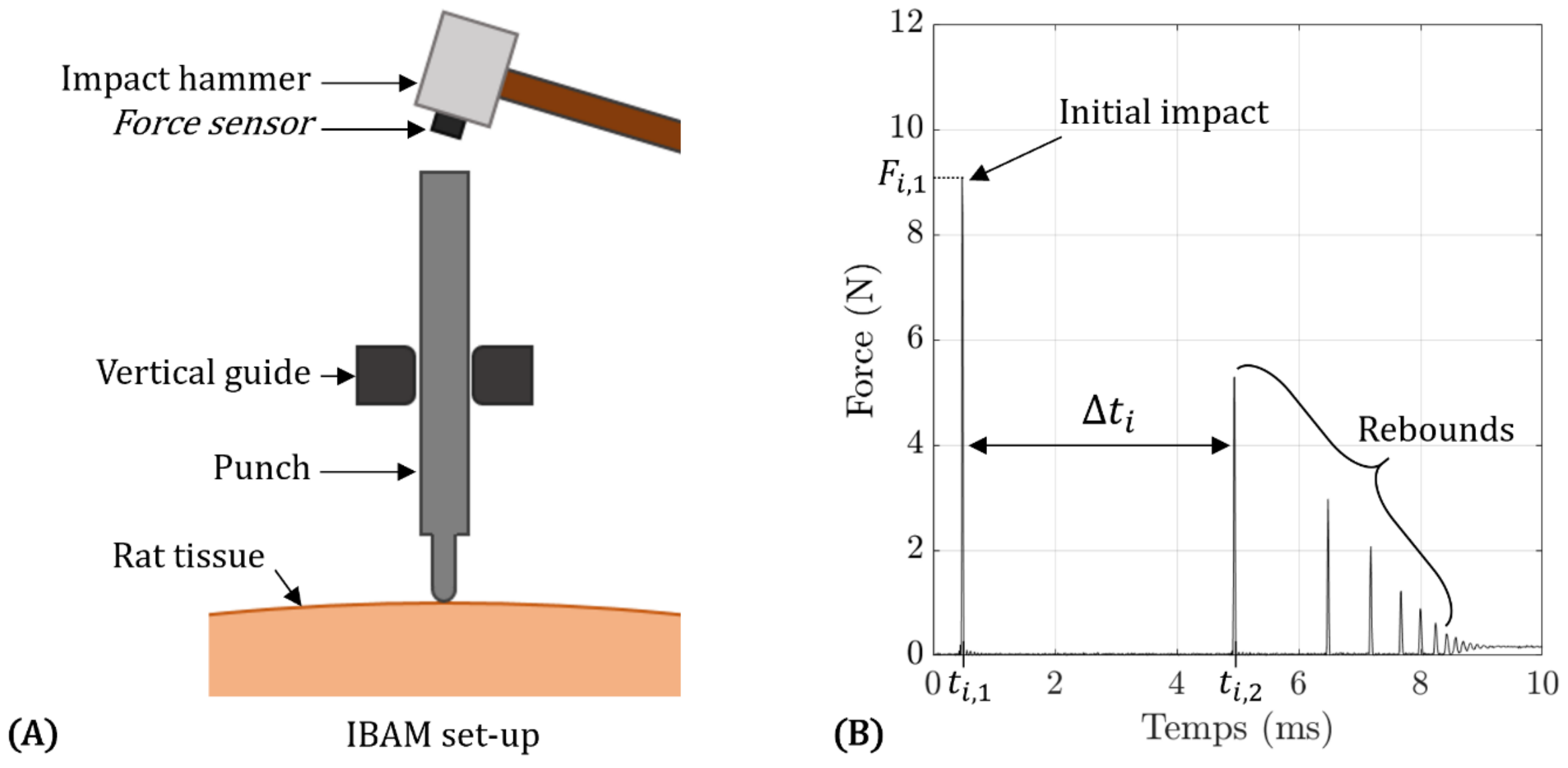

***Figure 3:*** ***(A)*** *Illustration of the experimental set-up of Impact-Based Analysis Method (IBAM).* ***(B)*** *Example of force signal obtained following hammer impact with* $\boldsymbol{\Delta t_i}$ *= 4.47 ms and* $\boldsymbol{F_{i,1}}$ *= 9.1 N.*

**Cutometer®** (MPA 580, Courage & Khazaka Electronic GmbH, Cologne, Germany) is a reference device in the cosmetic field for the assessment of the viscoelastic behavior of the skin (Dobrev, 2005; Serup et al., 2006; Ryu et al., 2008). The principle of this suction method is to i) exert a controlled negative pressure in order to deform the skin through a 2 mm-diameter aperture of the probe, and ii) record the skin elevation displacement by an optical measurement. Based on the curve of the tissue displacement as a function of time, several indicators are computed to reflect the mechanical properties of the superficial tissue. Here, for the purpose of simplification, only indicators $R0$ and $R5$ have been considered. According to the manufacturer, $R0$ refers to the passive behavior of superficial tissue under suction force and is defined as the tissue deformation at the end of the aspiration cycle (Serup et al., 2006), as shown in Fig. 4B. $R5$ corresponds to the net elasticity of the superficial tissue and is defined as the ratio of the elastic deformation during the relaxation phase $U_r$ and of the elastic deformation during the aspiration phase $U_e$ (Serup et al., 2006), as shown in Fig. 4B. In this protocol, a measurement was composed of 3 successive cycles of 2 s-aspiration and 2 s-relaxation with a negative pressure of 450 mbar (Cléry-Barraud et al., 2013; Nedelec et al., 2016; Kim et al., 2018) to provide a $R0$ and $R5$ mean value for each site, each rat and each experimental day.

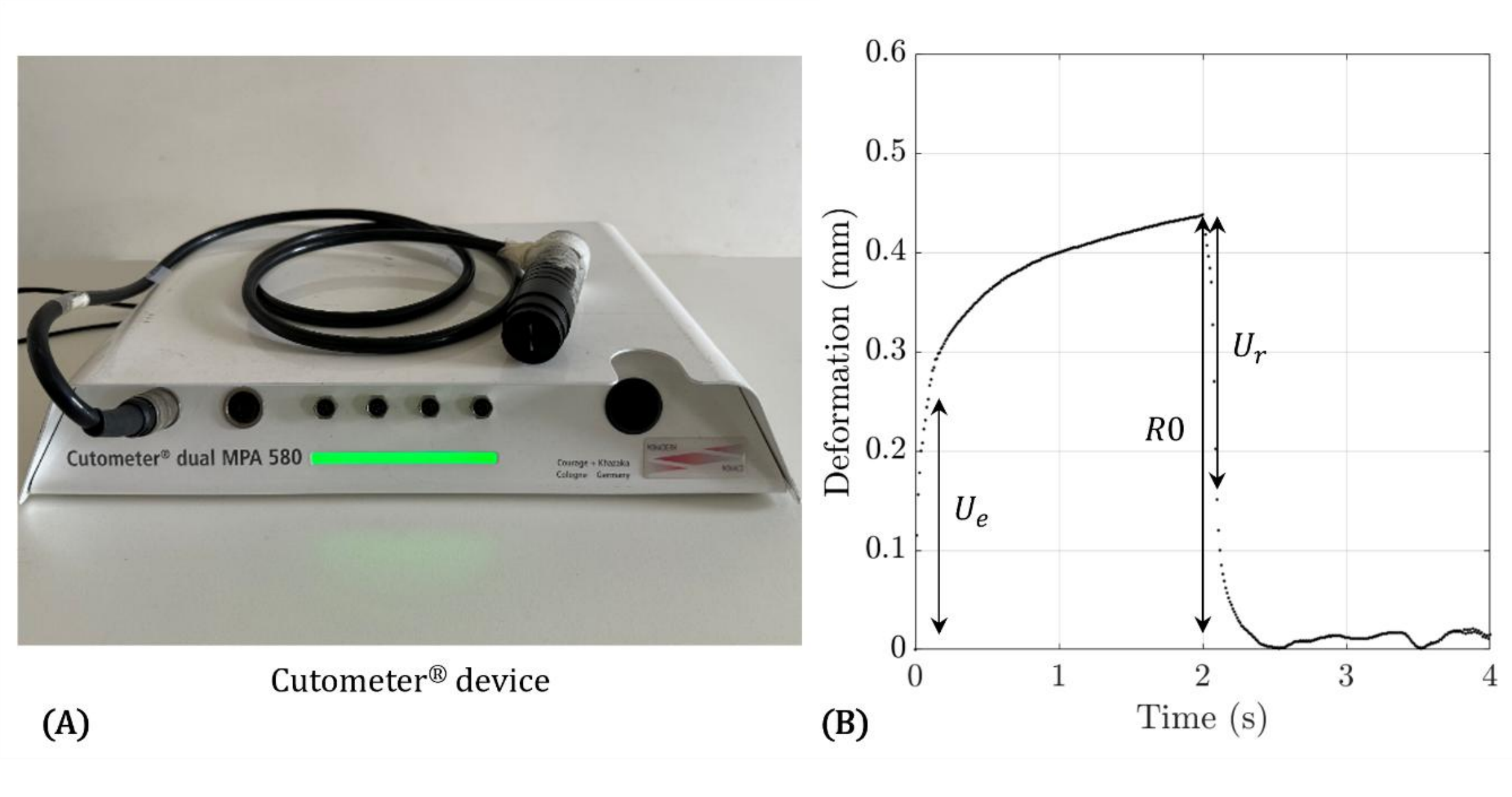

***Figure 4:*** ***(A)*** *Illustration of Cutometer® device with a 2 mm-diameter probe.* ***(B)*** *Example of tissue deformation during one cycle (aspiration / relaxation) with* $\boldsymbol{R0}$ *= 0.438 mm and* $\boldsymbol{R5} = \boldsymbol{U_r/U_e}$ *= 1.079.*

**MyotonPro®** (MyotonAS, Tallinn, Estonia) is a hand-held research device widely used for the non-invasive characterization of the biomechanical properties of various soft tissues (Bizzini and Mannion, 2003) such as muscles (Feng et al., 2018; Kelly et al., 2018), tendons (Feng et al., 2018) and skin (Dellalana et al., 2019; John et al., 2023). To perform a measurement, the probe of the device is positioned perpendicular to the superficial tissue. A pre-compression force is then imposed, before the generation of a mechanical pulse. Finally, based on the tissue response recorded by an accelerometer, and assuming a damped oscillating system, five parameters are calculated: the oscillation frequency $F$ (associated with the tissue tension), the dynamic stiffness $S$ (reflecting the resistance of the tissue under a compression force), the logarithmic decrement $D$ (reflecting the tissue damping), the mechanical stress relaxation time $R$ and the ratio of relaxation and deformation time $C$ (related to the viscoelastic behavior of the tissue) (Myoton, 2022). Nevertheless, only the indicator $S$ was considered in this study and is defined by the device as the product of the probe mass $m_p$ and the vertical acceleration at the instant of maximum compression force $a_{max}$ divided by the probe displacement at the same instant $\Delta l$ (Bizzini and Mannion, 2003; Myoton, 2022), as shown in Fig. 5B. In this protocol, a measurement was composed of 5 repetitions to provide a mean value of $S$ for each site, each rat, and each day.

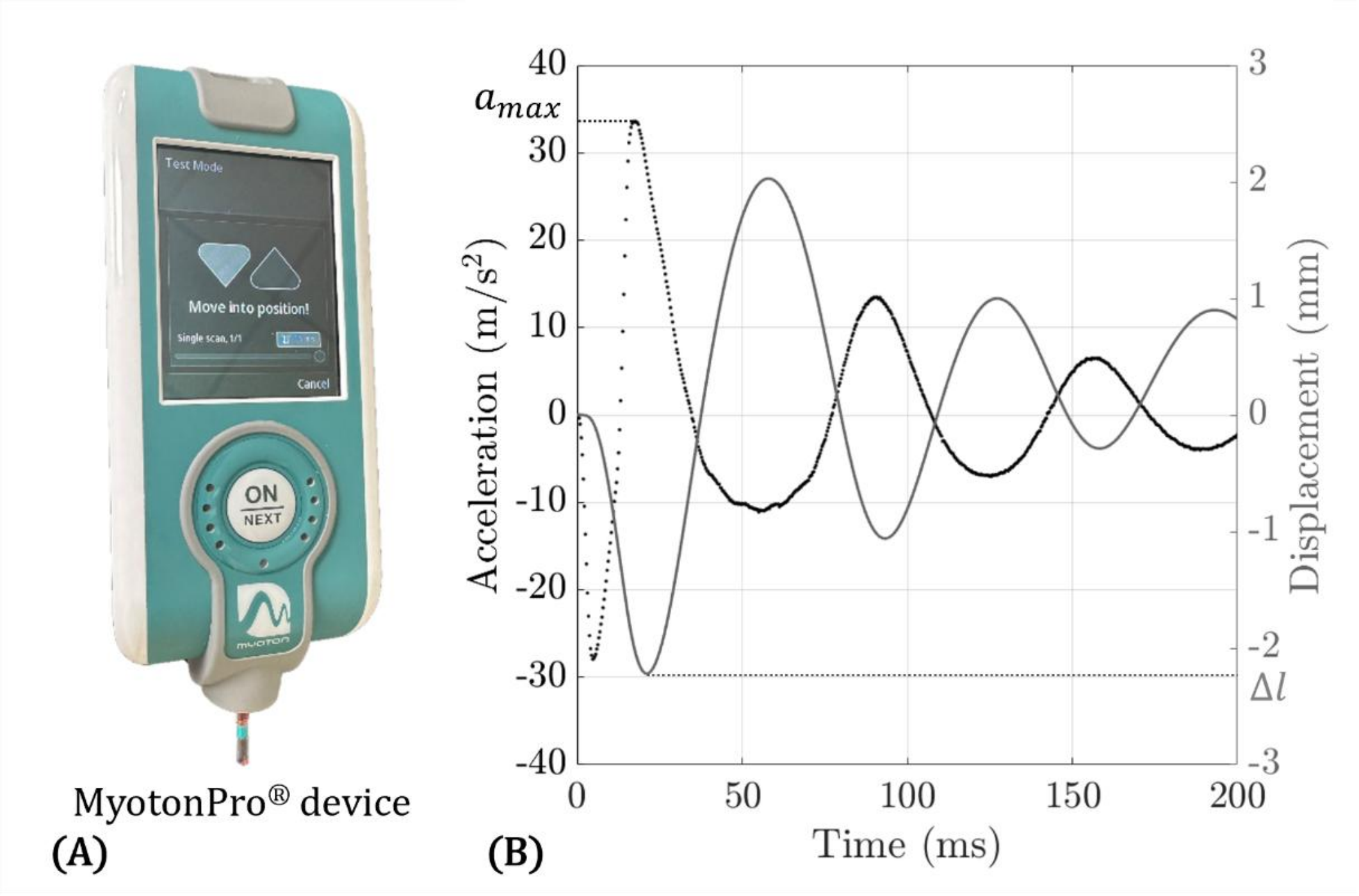

***Figure 5:*** ***(A)*** *Illustration of MyotonPro® device.* ***(B)*** *Example of acceleration and displacement of the tissue response as function of time with* $\boldsymbol{S = m_p \times a_{max}/\Delta l}$ *= 297 N/m.*

### 2.4 Data analysis

For each of the 4 indicators derived from the three mechanical surface characterization techniques ($\Delta t$ for IBAM, $S$ for MyotonPro® and $R0$ and $R5$ for Cutometer®), a dataset of 12 values (one value per rat) was obtained for each experimental day and for each region of interest. In order to remove inconsistent measurements, an outlier detection method using the z-score metric was applied to each dataset (Rousseeuw and Hubert, 2011; Schober et al., 2021). The threshold of z-score was fixed at 1.96 for a 95% confidence interval. The average and standard deviation values over the 12 rats of these corrected datasets were presented.

Prior to statistical analysis, the normality of the data for each dataset was checked using a Shapiro-Wilk test (Shapiro and Wilk, 1965). This procedure determined the form of the statistical hypothesis tests to be used: parametric if the data distribution follows a normal distribution, and non-parametric otherwise (Ghasemi and Zahediasl, 2012).

First, the effects of healing time on the 4 indicators were evaluated. To do so, each dataset obtained between $D_4$ and $D_{17}$ (post-wound generation or post-hyaluronic acid injection) was compared with the reference values of $D_0$ (pre-wound generation or pre-hyaluronic acid injection) using a two-sample test: Welch's t-test in the parametric case or Wilcoxon rank sum test in the non-parametric case. In both cases, the tests were performed with the left and right one-tailed options. The influence of healing time on the indicator values was also assessed using a one-way analysis of variance (ANOVA) in the parametric case, and a Kruskal-Wallis test in the non-parametric case. For all these statistical hypothesis tests, the significance level was fixed at 5%. When the results of the hypothesis tests were not significant according their p-value, the effect sizes were determined to check whether the lack of significance was due to low sensitivity or a true absence of effect. For a two-sample test, the effect size was interpreted using standard conventions (small for 0.2 < Cohen's d < 0.5, moderate for 0.5 < Cohen's d < 0.8, and large for Cohen's d ≥ 0.8) for a Welch's t-test or (small for $0.1 < r < 0.3$, moderate for $0.3 < r < 0.5$, and large for $r \geq 0.5$) for a Wilcoxon rank-sum test. For one-way ANOVA or Kruskal-Wallis test, the effect size was assigned as follows: small for $0.01 < \eta^2 < 0.06$, moderate for $0.06 < \eta^2 < 0.14$, and large for $\eta^2 \geq 0.14$.

Second, the values of the 4 indicators measured at the sites exposed to wound healing and hyaluronic acid injection and those at the control sites were compared using two approaches: statistical hypothesis tests (referred to as HT), and confidence intervals (CI), which allows to strengthen the interpretation of the results. The first approach (HT) used Welch's t-tests in the parametric case and Wilcoxon rank sum tests in the non-parametric case for the two-samples tests, similarly to the previous paragraph. For the statistical hypothesis testing (HT), the p-value only reflects the probability that the observed groups verify the null hypothesis, which corresponds to the situation where the two groups have equal means or medians for Welch's t-test and Wilcoxon rank sum test, respectively. In the second approach (CI), the differences between two groups are evaluated using a confidence interval of the mean difference, taking into account the relative reproducibility values of each characterization technique for the standard deviation. For the confidence interval of the mean difference (CI), the approach captures the behavior of the different possible scenarios. In this case, two groups are considered significantly different when the 95% confidence interval of the difference between the values of the two groups does not contain the null value. This CI approach usually allows to quantify the magnitude of the difference between two groups (Phillips et al., 2022), which is not the case with the HT approach that only allows to qualitatively compare the two groups. However, the HT approach is more often used in the literature. For both methods, the significance level was fixed at 5%. Using the same conventions mentioned in the previous paragraph, the effect sizes were estimated for hypothesis tests that showed no significance, based on their p-value. This was done to determine whether the lack of significance was due to low sensitivity or a true absence of effect.

# 3. Results

Throughout the protocol, the animals did not exhibit any deterioration in their physical condition and the affected skin regions recovered correctly. Nevertheless, one rat was lost during the study period because it did not wake up after the measurements of $D_7$.

## 3.1 Wound healing process

Figure 6 shows the evolution of the 4 indicators as a function of time during a wound healing process. For both "wound healing" and "control" groups and after $D_4$, $\Delta t$ (respectively, $R0$) values are lower (respectively, higher) than the reference values at $D_0$. For the "wound healing" group and after $D_4$, $R5$ values are higher than the reference values at $D_0$, while for the "control" group and after $D_{15}$, $R5$ values are higher than the reference values at $D_0$. In contrast, the values of $S$ after $D_4$ are not significantly different compared to the values at $D_0$, except for the values at $D_4$ in the "control" group. In the same way, the one-way statistical hypothesis tests confirm a significant influence of time on $\Delta t$, $R0$ and $R5$ for the "wound healing" and "control" groups. No significant effect of healing time was found for $S$, which reported small ($\eta^2$ = 0.041) and moderate ($\eta^2$ = 0.061) effect sizes for the "wound healing" and "control" groups, respectively. Note that for $D_4$ and $D_7$, it was not possible to use the Cutometer on the "wound healing" sites because of the presence of blood.

Table 1 summarizes the comparison between the "wound healing" and "control" groups at each healing time. The two data analysis approaches described in section 2.4 were applied to the data shown in Fig. 6. The values of $R5$ corresponding to the "wound healing" group are significantly higher than those of the "control" group during the entire measurement period ($D_{13}$ to $D_{17}$). For the indicator $R0$, the values of the "wound healing" group are lower than those for the "control" group at $D_{13}$ and $D_{15}$ with the statistical HT approach, while they are similar for the CI approach. A decrease of $\Delta t$ values between the "wound healing" and "control" groups is obtained at $D_4$, $D_7$ and $D_{15}$ (respectively $D_7$ and $D_{15}$) for the CI (respectively HT) approach. Eventually, there are no consistent behaviors between $S$ values, which is to be expected given the small and negligible effect sizes of hypothesis tests. However, this is not the case for $D_4$, where the values of the "wound healing" group are lower than those of the "control" group according to both data analysis approaches. Table 2 summarizes the experimental conditions in which significant indicator variations were noted using statistical methods.

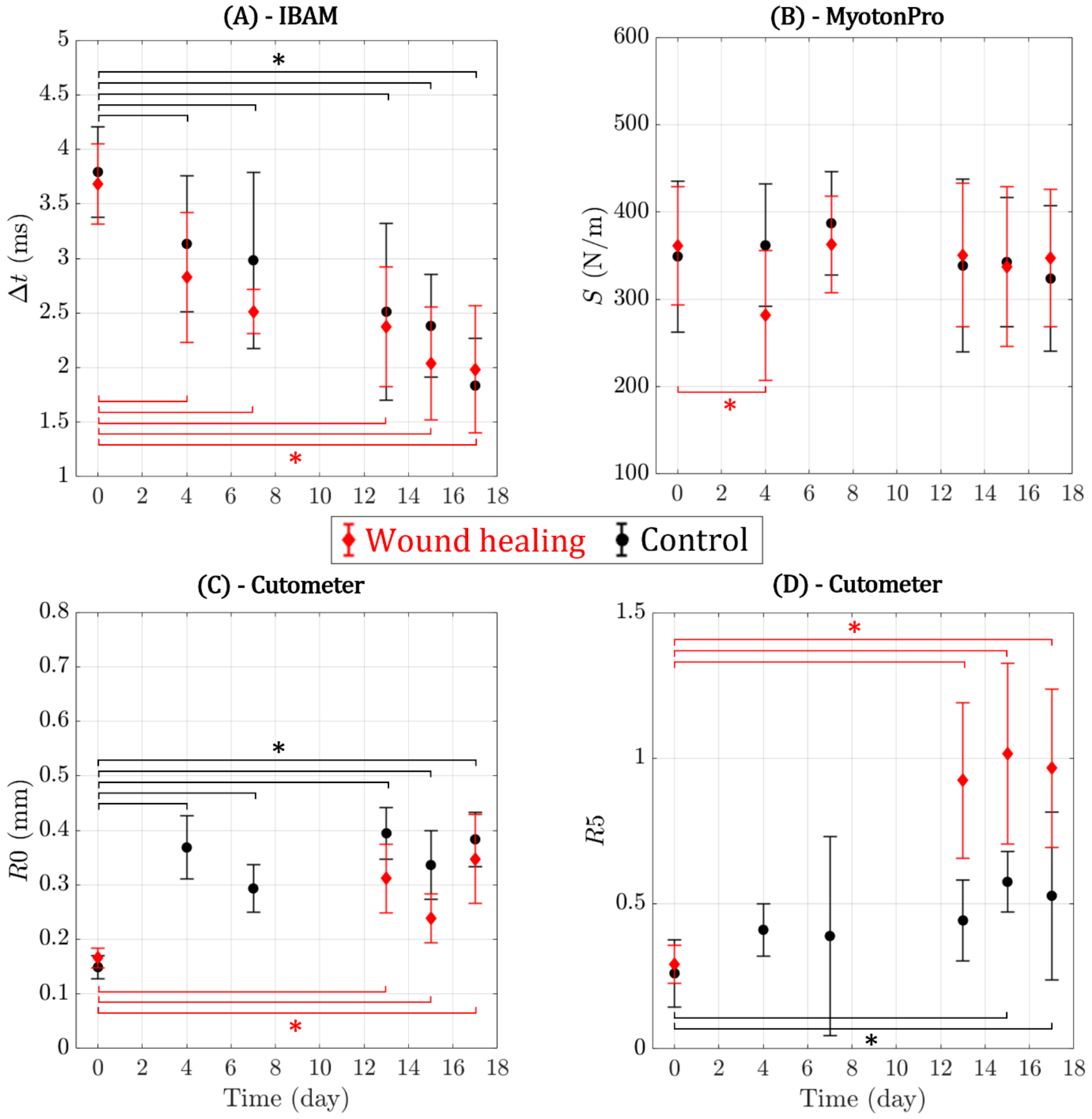

***Figure 6:*** *Variation of the different indicators for wound healing and control sites for the different healing times:* ***(A)*** $\boldsymbol{\Delta t}$ *obtained from IBAM,* ***(B)*** $\boldsymbol{S}$ *obtained from MyotonPro,* ***(C)*** $\boldsymbol{R0}$ *and* ***(D)*** $\boldsymbol{R5}$ *obtained from Cutometer. The values for day 0 were measured prior to wound generation event. Each point corresponds to the mean ± standard deviation. The significant differences (p-value < 0.05) with respect to day 0 are indicated by significance brackets with * (respectively *) for values from "wound healing" (respectively "control") groups.*

| Characterization techniques | Statistical analysis approaches | Pre- | Post-wound generation | | | | |
|---|---|---|---|---|---|---|---|
| | | $D_0$ | $D_4$ | $D_7$ | $D_{13}$ | $D_{15}$ | $D_{17}$ |
| IBAM - $\Delta t$ | *HT* | • | • | - | • | - | • |
| | *CI* | • | - | - | • | - | • |
| MyotonPro - $S$ | *HT* | • | - | • | • | • | • |
| | *CI* | + | - | - | + | - | + |
| Cutometer - $R0$ | *HT* | + | | | - | - | • |
| | *CI* | • | | | • | • | • |
| Cutometer - $R5$ | *HT* | • | | | + | + | + |
| | *CI* | • | | | + | + | + |

***Table 1:*** *Comparison between the "wound healing" and "control" groups using the statistical hypothesis tests (HT) and the confidence interval of the mean difference (CI). For both approaches, the sign + (respectively, -) indicates that the values obtained for the "wound healing" group are significantly higher (respectively, lower) than the values obtained for the "control" group at a level of 95%. • indicates no significance between the two groups.*

| Characterization techniques | Statistical analysis approaches | Wound healing **vs** Control |
|---|---|---|
| IBAM - $\Delta t$ | *HT* | **Significant effect** |
| | *CI* | **Significant effect** |
| MyotonPro - $S$ | *HT* | No trends |
| | *CI* | No trends |
| Cutometer - $R0$ | *HT* | **Significant effect** |
| | *CI* | No trends |
| Cutometer - $R5$ | *HT* | **Significant effect** |
| | *CI* | **Significant effect** |

***Table 2:*** *Summary of the statistical analysis results between the "wound healing" and "control" groups using the statistical hypothesis tests (HT) and the confidence interval of the mean difference (CI).*

### 3.2 Hyaluronic acid injection

Figure 7 shows the variation of the 4 indicators as a function of healing time after an intradermal hyaluronic acid injection. For "HA – 0.5mL" groups and after $D_4$, the values of $R0$ and $R5$ are significantly higher than the reference values at $D_0$, except at $D_{13}$ with $R5$. Only after $D_{13}$, the values of $\Delta t$ are lower than the reference values at $D_0$. In contrast, for both "HA – 0.25mL" and "control" groups and after $D_4$, $\Delta t$ (respectively, $R0$ and $R5$) values are significantly lower (respectively higher) than the reference values at $D_0$. For the "control" group, at $D_4$, $D_7$ and $D_{17}$, the values of $S$ are higher than the values at $D_0$. No other significant variation was found for $S$, except for the "HA - 0.25mL" group at $D_{15}$, where the value of $S$ is lower than the value at $D_0$. Similarly, a one-way statistical hypothesis tests confirm a significant influence of time on the $\Delta t$, $R0$ and $R5$ indicator values for the “HA – 0.25mL”, “HA – 0.5mL” and “control” groups, while it is not the case for the values of $S$, despite moderate ($\eta^2$ = 0.074 and $\eta^2$ = 0.106) and large ($\eta^2$ = 0.150) effect sizes for the “control”, “HA – 0.25mL” and “HA – 0.5mL” groups, respectively.

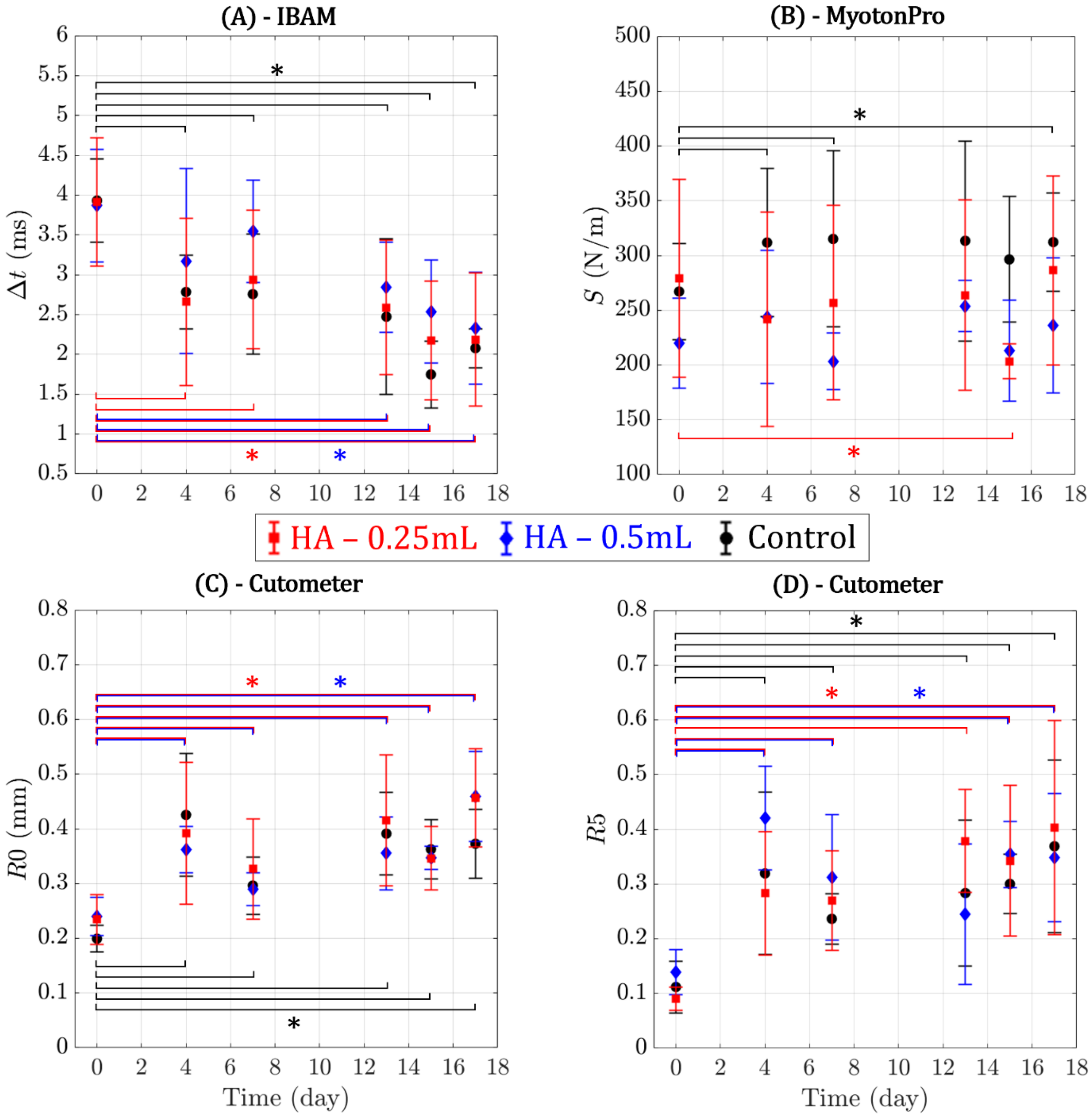

***Figure 7:*** *Variation of the mechanical parameters of intradermal hyaluronic acid injection (0.5 or 0.25 mL dose) and control sites on the lower back over time:* ***(A)*** $\boldsymbol{\Delta t}$ *from IBAM,* ***(B)*** $\mathbf{S}$ *from MyotonPro,* ***(C)*** $\mathbf{R0}$ *and* ***(D)*** $\mathbf{R5}$ *from Cutometer. The values for day 0 were measured prior to hyaluronic acid injection. Each point corresponds to the mean ± standard deviation. The significant differences (p-value < 0.05) with respect to day 0 are indicated by significance brackets with * (respectively * and *) for values from “HA – 0.5 mL” (respectively “HA – 0.25 mL” and “control”) groups.*

| Characterization techniques | Statistical analysis approaches | | Pre- | Post-hyaluronic acid injection | | | | |
|---|---|---|---|---|---|---|---|---|
| | | | $D_0$ | $D_4$ | $D_7$ | $D_{13}$ | $D_{15}$ | $D_{17}$ |
| IBAM - $\Delta t$ | *HT* | HA - 0.5mL vs Control | • | • | + | • | + | • |
| | *CI* | | • | + | + | + | + | • |
| MyotonPro - $S$ | *HT* | | - | - | - | - | - | - |
| | *CI* | | - | - | - | - | - | - |
| Cutometer - $R0$ | *HT* | | + | • | • | • | • | + |
| | *CI* | | • | • | • | • | • | • |
| Cutometer - $R5$ | *HT* | | • | + | • | • | • | • |
| | *CI* | | • | • | • | • | • | • |
| IBAM - $\Delta t$ | *HT* | HA - 0.25mL vs Control | • | • | • | • | • | • |
| | *CI* | | • | • | • | • | + | • |
| MyotonPro - $S$ | *HT* | | • | • | • | • | - | • |
| | *CI* | | + | - | - | - | - | - |
| Cutometer - $R0$ | *HT* | | + | • | • | • | • | + |
| | *CI* | | • | • | • | • | • | • |
| Cutometer - $R5$ | *HT* | | • | • | • | + | • | • |
| | *CI* | | • | • | • | • | • | • |
| IBAM - $\Delta t$ | *HT* | HA - 0.5mL vs HA - 0.25mL | • | • | • | • | • | • |
| | *CI* | | • | + | + | • | + | • |
| MyotonPro - $S$ | *HT* | | • | • | • | • | • | • |
| | *CI* | | - | • | - | - | + | - |
| Cutometer - $R0$ | *HT* | | • | • | • | • | • | • |
| | *CI* | | • | • | • | • | • | • |
| Cutometer - $R5$ | *HT* | | + | + | • | - | • | • |
| | *CI* | | • | • | • | • | • | • |

***Table 3****: Results of the statistical analysis between the "HA – 0.5mL", "HA – 0.25mL" and "control" groups using the statistical hypothesis tests (HT) and the confidence interval of the mean difference (CI). For each comparison, the sign + (respectively, - ) indicates that the values obtained for the first group are significantly higher (respectively, lower) than for the second group at a level of 95%. • indicates no significance between the two groups.*

Table 3 summarizes the differences obtained between the "HA – 0.25mL", "HA – 0.5mL" and "control" groups for the healing times. The same two data analysis approaches as in subsection 3.1 were applied to the data shown in Figure 7. According to both approaches, the values of $S$ for the "HA – 0.5mL" group are significantly lower than those for the "control" group after $D_4$. Moreover, according to the CI approach, the values of $S$ for the "HA - 0.25mL" group are significantly lower than those of the "control" group after $D_4$, which is confirmed by the HT approach for $D_{15}$ only. Similarly, according to the CI approach, the values of $S$ for the "HA - 0.5mL" group are significantly lower than those of the "HA – 0.25mL" group at $D_7$, $D_{13}$ and $D_{17}$. Similarly, the values of $\Delta t$ for the "HA - 0.5mL" group are significantly higher than those for the "control" group between $D_4$ and $D_{15}$ using the CI approach, which is confirmed with the HT approach at $D_7$ and $D_{15}$. The values of $\Delta t$ for the "HA - 0.5mL" group are also significantly higher than those of the "HA - 0.25mL" group at $D_4$, $D_7$ and $D_{15}$ according to the CI approach. Eventually, the two data analysis approaches reveal no clear tendencies with $R0$ or $R5$ values between the different groups, which is to be expected given the small and moderate effect sizes of hypothesis tests. Table 4 summarizes the experimental conditions in which significant indicator variations were noted using statistical methods.

| Characterization techniques | Statistical analysis approaches | HA - 0.5mL **vs** Control | HA - 0.25mL **vs** Control | HA - 0.5mL **vs** HA - 0.25mL |
|---|---|---|---|---|
| IBAM - $\Delta t$ | *HT* | **Significant effect** | No trends | No trends |
| | *CI* | **Significant effect** | No trends | **Significant effect** |
| MyotonPro - $S$ | *HT* | **Significant effect** | No trends | No trends |
| | *CI* | **Significant effect** | **Significant effect** | **Significant effect** |
| Cutometer - $R0$ | *HT* | No trends | No trends | No trends |
| | *CI* | No trends | No trends | No trends |
| Cutometer - $R5$ | *HT* | No trends | No trends | No trends |
| | *CI* | No trends | No trends | No trends |

***Table 4:*** *Summary of the statistical analysis results between the "HA – 0.5mL", "HA – 0.25mL" and "control" groups using the statistical hypothesis tests (HT) and the confidence interval of the mean difference (CI).*

### 3.3 Control region

Figure 8 shows the evolution of the 4 indicators as a function of time in the healthy median ventral region. As in subsection 3.1, after $D_4$, $\Delta t$ (respectively, $R0$) values are lower (respectively, higher) than the reference values at $D_0$, except for $D_7$. In contrast, the values of $S$ (respectively, $R5$) increase compared to the values at $D_0$ only for $D_4$ and $D_7$ (respectively, $D_{15}$ and $D_{17}$). The one-way statistical hypothesis tests reported a significant influence of time on the $\Delta t$ and $R0$ indicator values, while the effect was not found significant for $R5$ and $S$, despite moderate ($\eta^2$ = 0.070) and large ($\eta^2$ = 0.149) effect sizes, respectively.

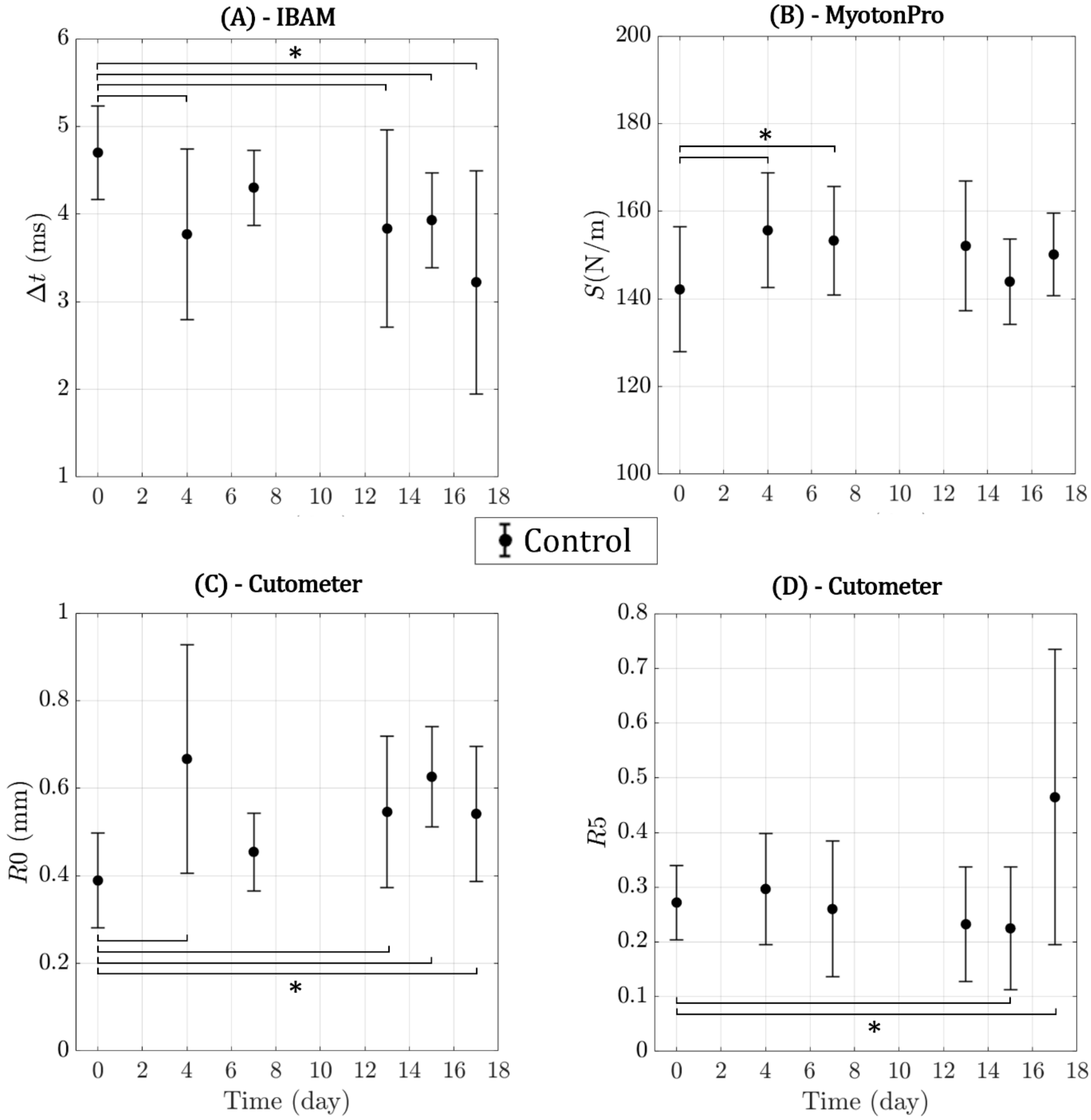

***Figure 8:** Variation of the mechanical parameters of median ventral control site as a function of time: **(A)** $\boldsymbol{\Delta t}$ from IBAM, **(B)** $\boldsymbol{S}$ from MyotonPro, **(C)** $\boldsymbol{R0}$ and **(D)** $\boldsymbol{R5}$ from Cutometer. Each point corresponds to the mean ± standard deviation. The significant differences (p-value < 0.05) with respect to day 0 are indicated by significance brackets with *.*

### 3.4 Repeatability and Reproducibility

Based on the measurement described in the last paragraph of subsection 2.2, Table 5 reports the average repeatability and reproducibility values obtained at the different sites (the left and right upper back sites and the median ventral site) for the four indicators. The lowest repeatability values are obtained with the indicators $S$ and $\Delta t$ (respectively, 3.2% and 3.5%). The indicator $S$ has the best reproducibility.

| Characterization techniques | *Repeatability* | *Reproducibility* |
|---|---|---|
| IBAM - $\Delta t$ | 3.5 % | 12.6 % |
| MyotonPro - $S$ | 3.2 % | 6.2 % |
| Cutometer - $R0$ | 8.4 % | 11.8 % |
| Cutometer - $R5$ | 13.2 % | 28.0 % |

***Table 5:*** *Results of average relative repeatabilities and reproducibilities of* $\boldsymbol{\Delta t}$ *from IBAM,* $\boldsymbol{S}$ *from MyotonPro,* $\boldsymbol{R0}$ *and R5 from Cutometer. These values were estimated from 5 repetitions in 3 different regions.*

# 4. Discussion

The originality of this *in vivo* study was to characterize the biomechanical properties of superficial tissues using three different non-invasive techniques, in relation with two physiological processes: wound healing after full-thickness excisional wound and skin regeneration after intradermal injection of hyaluronic acid (HA). Another originality was to apply the IBAM for the first time *in vivo*. The objective was also to confront and validate a new mechanical surface characterization approach, IBAM, through measurements performed *in vivo* and to compare its performance with two other devices: Cutometer and MyotonPro.

## 4.1 Non-invasive mechanical surface characterization techniques

**Cutometer** is a reference technique in the cosmetic industry for the measurement of the viscoelastic properties of human skin. The device exhibits some reproducibility issues, primarily because of how the probe is held against the skin (Stroumza et al., 2015). The variability is primarily caused by the perpendicularity and load exerted by the probe on the tissue (Serup et al., 2006; Bonaparte et al., 2013; Stroumza et al., 2015), which constitutes a significant source of manipulation bias (Stroumza et al., 2015). The Cutometer has also widely been used with animal models such as murine models in order to investigate the behavior of tissue mechanical properties during the wound healing (Cléry-Barraud et al., 2013; Held et al., 2015) or rejuvenation (Heo et al., 2021) processes of the skin. In the same context, Cutometer was also used to quantify the biomechanical effects of various treatments, such as the choice of dressing or wound healing treatments (Held et al., 2016; Lintel et al., 2023), skin rehydration products (Seok et al., 2016) or skin rejuvenation methods (Hong et al., 2021; Ghasemi et al., 2023; Charipoor et al., 2024). Since the different indicators obtained with the Cutometer are highly interconnected (Draaijers et al., 2004), only the $R0$ and $R5$ indicators were selected herein for the sake of simplicity. $R0$ is one of the most widely used Cutometer indicators, along with $R2$ (Abbas et al., 2022). However, these two parameters are dependent on tissue thickness (Serup et al., 2006; Abbas et al., 2022). $R0$ and $R2$ depend on both the stiffness and thickness of superficial tissues, which should be taken into account in the interpretations of the results. Skin thickness is known to depend on age and on the anatomical location. In contrast, the dimensionless indicators: $R5$, $R6$ and $R7$ are not affected by tissue thickness. Nevertheless, among these three indicators, $R5$ was the only one capable of significantly differentiating between healthy skin and fibrotic scars (Abbas et al., 2022).

**MyotonPro** is a palpation device available for research applications and widely used on humans. The device has already proven to be reliable for assessing the viscoelastic properties of skin (Dellalana et al., 2019; Gilbert et al., 2021). However, very few studies have used MyotonPro to characterize the biomechanical properties of skeletal muscle in rats (Alev et al., 2018, 2022) or tendons in pig models (Anggoro et al., 2024) or dogs (Kurashina et al., 2023). In fact, MyotonPro is primarily intended for measuring muscle and ligament stiffness (Rosicka et al., 2021). The positioning of the device relative to the tissue can alter tissue measurements, because of contributions from subcutaneous tissues such as muscles or bones (Dellalana et al., 2019). Additionally, Bartsch et al. (Bartsch et al., 2023) demonstrated that MyotonPro may not be able to measure thin layers in multilayered structures such as the skin. Since according to the manufacturer, the other parameters obtained with the MyotonPro are related to viscosity or surface tension, only the indicator $S$ was considered herein.

**IBAM** is a more recent set-up developed by our group. Based on modeling (Poudrel et al., 2024) and on experimental data (Bouffandeau et al., 2025), previous works have shown that $\Delta t$ is roughly inversely proportional to the square root of the tissue Young's modulus (Poudrel et al., 2024; Bouffandeau et al., 2025). Furthermore, in a previous *in vitro* comparative study (Bouffandeau et al., 2026), robust correlations were found between the Young's modulus of synthetic soft tissue phantoms and the indicators $\Delta t$, $S$ and $R0$. The following paragraph compares in more details the performances of the 3 devices used in the present study.

Table 6 shows a summary of the effect of tissue stiffness variations on the different indicators, which was derived from (Bouffandeau et al., 2026).

| Characterization techniques | **Effects of Increasing Tissue Stiffness** |
|---|---|
| IBAM - $\Delta t$ | ⮟ *(decrease)* |
| MyotonPro - $S$ | ⮝ *(increase)* |
| Cutometer - $R0$ | ⮟ *(decrease)* |
| Cutometer - $R5$ | ⮝ *(increase)* |

***Table 6:*** *Effect of increasing tissue stiffness on the different indicators. Derived from (Bouffandeau et al., 2026).*

### 4.2 Biomechanical characterization

**Age dependence.** For all the regions of interest, time was shown to significantly influence the variation of the indicators: $\Delta t$, $R0$ and $R5$ (except for the median ventral site with $R5$). The value of the indicator $\Delta t$ (respectively, $R0$ and $R5$) was shown to decrease (respectively, increase) as function of time, with a more pronounced effect on the back than on the ventral sites. Thus, the variations of $\Delta t$ and $R0$ indicators are opposite. However, Bouffandeau et al. (Bouffandeau et al., 2026) demonstrated that the values of these two indicators decreased with the Young's modulus of the characterized material. The rats were 7 weeks old at $D_0$. Significant evolution of the morphological and biomechanical characteristics of Sprague-Dawley rats are known to occur between 7 and 10 weeks. Between the 1$^{st}$ and 4$^{th}$ months, the weight of the rats is multiplied by 4 (from approx. 100 g to 400 g) (Vogel, 1978, 1981). Moreover, the thickness of their back skin is multiplied by 2 (from 0.5 mm to 1.6 mm between 1$^{st}$ and 4$^{th}$ month) (Vogel, 1978, 1981), mainly because of a significant increase of the dermis thickness (from 200 µm to 1000 µm between 3$^{rd}$ week and 2$^{nd}$ month) (Kazanci et al., 2017). The rat weight and skin thickness are known to reach their maximum values around the 12$^{th}$ month (Vogel, 1981). Over the same period, the modulus of elasticity in traction also quadruples (from around 10 to 40 MPa) (Vogel, 1978) between the 1st and 4th months. All the aforementioned physiological changes can explain the relative variations obtained for the indicators $\Delta t$, $R0$ and $R5$. However, time was shown not to influence the variation of the indicator $S$. Nevertheless, given the moderate and large effect sizes observed in the hypothesis tests, this observation may suggest a lack of sensitivity of the indicator $S$ of the MyotonPro to the properties of the superficial soft tissue. It is reasonable to assume that subcutaneous tissue represents the majority of overall tissue measurements, rather than skin variations when using the MyotonPro. According to the manufacturer (Myoton, 2022), the MyotonPro has a measurement depth of around 20 mm, which has been verified in a previous study using soft tissue phantoms (Bouffandeau et al., 2025). This value is consistent with the main application of the device which corresponds to the mechanical characterization of muscles and tendons (Rosicka et al., 2021).

Finally, IBAM and Cutometer appear to detect significant age-related variations in the mechanical behavior of superficial soft tissue in all measurement regions, in contrast to MyotonPro.

**Wound healing process.** IBAM and Cutometer demonstrated could assess significant differences of the mechanical behavior of the “wound healing” and “control” groups. The values of the indicators $\Delta t$ and $R0$ (respectively, $R5$) obtained for the “wound healing” group are lower (respectively, higher) compared to the ones obtained for the healthy tissue until $D_{15}$ (respectively, until $D_{17}$). In contrast, no significant variation was observed between the two groups when considering the indicator $S$. Similar results have been obtained in other studies considering the variation of the mechanical properties of superficial tissues during wound healing (Cléry-Barraud et al., 2013; Held et al., 2015) using the Cutometer on murine models. Working with SKH-1 male mice, Cléry-Barraud et al. (Cléry-Barraud et al., 2013) reported significant differences of the values of several Cutometer indicators ($R2$, $R5$, $R6$ and $R7$) from $D_{14}$ to $D_{21}$ between the measurements of healthy skin and scars induced by sulphur mustard exposure. However, this study did not consider $R0$ due to its dependence on tissue thickness. Similarly to our results, the values of R5 were higher for the “wound healing” group compared to “control” group. Moreover, working with Lewis adult male rats, Held et al. (Held et al., 2015) showed a non-significant decrease of $R0$ values during the first 30 days in regions with partial-thickness wounds (up to 0.3 mm in depth, encompassing the epidermis and papillary dermis). However, this study (Held et al., 2015) used a Cutometer probe with an internal diameter of 6 mm which is different compared to the present study. Eventually, Draaijers et al. (Draaijers et al., 2004) suggested that the scar elastic properties could be objectively assessed using the Cutometer. Indeed, they demonstrated the correlation between i) $R0$ and $R5$ and ii) the subjective evaluation of scar pliability, using a scale ranging from 1 (i.e. normal skin) to 10 (the stiffest conceivable scar), which is commonly used in medical practice.

Finally, IBAM and Cutometer can quantitatively monitor variations of the biomechanical properties of superficial soft tissue during the wound healing process, in contrast to MyotonPro. In fact, the indicator $\Delta t$ for IBAM, and $R0$ and $R5$ for Cutometer reveal a significant increase in tissue stiffness in wound healing-affected regions compared to healthy tissue.

**Regeneration process.** IBAM and MyotonPro could be used to assess the changes of the mechanical properties of superficial tissues following intradermal injection of hyaluronic acid, at both lower back sites. The values of $S$ (respectively, $\Delta t$) are lower (respectively, higher) in the "HA - 0.5mL" and "HA - 0.25mL" groups compared to the “control” group for all post-injection days (respectively, from $D_4$ to $D_{15}$). Moreover, the values of $S$ (respectively, $\Delta t$) were lower (respectively, higher) in the "HA - 0.5mL" group than in the "HA - 0.25mL" group.

However, for $S$ and $\Delta t$, the differences between the "HA - 0.5mL" and "HA - 0.25mL" groups were less important than in comparisons with the “control” group. Conversely, the indicators obtained with the Cutometer showed no global significant difference between the "HA” and “control” groups, even though isolated variations could be observed, such as an increase of the values of $R5$ in the "HA - 0.5mL" group compared to the "HA - 0.25mL" and “control” groups at $D_4$ or an increase of $R0$ values in the "HA - 0.5mL" and "HA - 0.25mL" groups compared to the “control group” from $D_{17}$. Note that at $D_4$, the difference of the values of $R5$ between the “HA” and “control” groups could be due to the surface tension induced by the spherical-shaped nodule caused by the injection of hyaluronic acid, which then progressively disappears over time. This nodule is caused by the hydrophilic nature of HA that generates immediate volume filling after injection (Gilbert et al., 2012).

After $D_4$, the results obtained in the present study with MyotonPro and IBAM demonstrate a decrease of the stiffness of superficial tissues in response to the injection of HA over the first weeks. However, according to the literature (Gilbert et al., 2012), the effects of HA should be present for up to 6 to 12 months after injection due to its assimilation within the reticular dermis (Tran et al., 2014). This apparent contradiction might be explained by the fact that IBAM may primarily detect superficial changes rather than those associated with HA integration within deeper dermal layers, particularly the reticular dermis. In contrast, IBAM could be less sensitive to the medium and long-term effects of HA (tissue hydration retention) (Manuskiatti and Maibach, 1996; Papakonstantinou et al., 2012), which affect the deeper layer of the skin (i.e. reticular dermis). Increasing the measurement depth of IBAM could potentially improve its sensitivity to the medium and long-term effects of HA. However, it would be more interesting to investigate how the mechanical properties of the different layers of the characterized tissues contribute to indicator $\Delta t$, as well as to the recorded force signal. In a recent study performed with 6 week-old female Sprague-Dawley rats, Hong et al. (Hong et al., 2021) found an increase of $R5$ only from 2 weeks after the injection of another type of skin rejuvenation product: polycaprolactone. Finally, it may also be necessary to consider longer studies to validate the sensitivity of IBAM to the medium- and long-term effects of HA.

Finally, IBAM and MyotonPro can quantitatively monitor variations of the biomechanical properties of superficial soft tissue following intradermal injection of hyaluronic acid, in contrast to Cutometer. In fact, the indicator $\Delta t$ for IBAM and $S$ for MyotonPro reveal a significant decrease of tissue stiffness affected by hyaluronic acid injection compared to healthy tissue.

### 4.3 Limitations and perspectives

Firstly, an important limitation lies in the choice of animal model to mimic the mechanical behavior of human skin. The procurement and handling of this type of animal is low-cost, accessible and simple compared to other animals (Cléry-Barraud et al., 2013; Zomer and Trentin, 2018). Murine models have a similar skin structure compared to humans, with three layers: epidermis, dermis and hypodermis (Zomer and Trentin, 2018). However, there are significant differences between human and murine skins. The skin thickness is significantly lower for murine models (less than 25 µm) compared to that of humans (higher than 1 mm) (Zomer and Trentin, 2018). Furthermore, murine skin has an extra-muscular layer located in the subcutaneous tissue, called panniculus carnosus, which has a strong influence on the wound healing process (Zomer and Trentin, 2018). Porcine models could be a more satisfactory option for future studies (Lindblad, 2008) because of a morphology and skin structure closer to human. Additionally, this animal model could be used to study keloids or hypertrophic scars (Lindblad, 2008), which are difficult to reproduce in other animals.

Secondly, the experimental protocol has various limitations. The wounds healing process occurs in four successive phases (hemostasis, inflammation, proliferation and remodeling) (Wallace et al., 2026). This physiological mechanism is slightly faster in rats than in humans (Vidinský et al., 2006). According to histological analyses following skin incisions in Sprague-Dawley rats, the inflammatory phase occurs during the first three days, the proliferation phase reaches its maximal intensity around the fifth and sixth days, and the remodeling phase begins on the seventh day and can last several months (Vidinský et al., 2006). In light of the speed of the physiological mechanisms, the measurement frequency could have been increased. Furthermore, extending the study duration could be relevant to refine the monitoring of the end of the wound healing process, *i.e.* the remodeling phase. During the first two weeks after the lesion generation, measurements should have been taken every day and then every 2-3 days over a period of 2-3 months. Moreover, during the first week, the presence of blood and the incomplete scar closure affected the measurements made with the Cutometer. The presence of liquid or a scab at the measurement site can disrupt optical measurements or obstruct tissue deformation. Consequently, no measurements are conducted under these conditions, similarly as was done in Cléry-Barraud et al. (Cléry-Barraud et al., 2013). Thus, less invasive approaches than full-thickness excisional wound could be a source of improvement to minimize the formation of scab or cavity. For example, a dermatome could be used to reduce the full-thickness excisional wound to a partial-thickness excisional wound with a depth of 0.3 mm (i.e. only the epidermis and papillary dermis), while limiting the potential risks for the animal (Rahmanian-Schwarz, et al. 2011; Held et al., 2015). Meanwhile, various parameters used for the 3 measurement techniques could be modified. Regarding the MyotonPro, the use of another probe (e.g. the L-shape probe), could offer a more superficial characterization more suited to skin characterization (Rosicka et al., 2021). Nevertheless, using such probe would lead to the characterization of the shear behavior of superficial soft tissues, which would be more difficult to compare with the results obtained with IBAM and Cutometer (Junker et al., 2023). Regarding the Cutometer, other probes are also available with interior diameters of 6 and 8 mm, allowing a deeper characterization. With human skin, the 2 mm diameter is designed to evaluate the epidermal layer, while the 6 mm diameter is designed to measure the dermal layer (Junker et al., 2023).

Thirdly, the IBAM is still under development. Some technical choices were changed compared to previous versions of the experimental set-up implemented for soft tissue phantoms (Bouffandeau et al., 2025, 2026). In particular, the geometry of the bottom of the punch, which corresponds to the region in contact with the tissue, has been changed. The diameter of this region has been reduced from 4 to 2 mm to reduce the depth of measurement (Bouffandeau et al., 2025) and the shape of the punch end has been changed to a half-sphere to minimize variations in reproducibility as a function of the angulation of the punch with the tissue.

## 5. Conclusion

The purpose of this paper was to quantitatively monitor the biomechanical behavior of the skin of a murine model during wound healing and after an intradermal injection of hyaluronic acid. To do so, three techniques (simple, low-cost, objective and non-invasive) were used, including our Impact-Based Analysis Method (IBAM) and 2 existing devices: Cutometer and MyotonPro. This study is the first one to use IBAM *in vivo*. The changes of the mechanical properties of superficial soft tissues were estimated *in vivo* in various healthy and affected regions over a 3-week period using the indicators $\Delta t$ for IBAM, $S$ for MyotonPro, $R0$ and $R5$ for Cutometer. Compared to healthy control regions, the couple of indicators $\Delta t$ and $R0$ (respectively, $\Delta t$ and $S$) reported an increasing (respectively, decreasing) tissue stiffness in the wound healing (respectively, hyaluronic acid-injected) regions, respectively. IBAM was the only technique that showed significant variations in mechanical properties with respect to both mechanisms. Additionally, IBAM and Cutometer were sensitive to the animal age. Further studies should i) consider an animal model more consistent with the human model in terms of skin structure, ii) extend the healing period and reduce the measurement intervals of the protocol. The present study paves the way for the development of a portable, easy-to-use device using IBAM. After automating the measurement process, an IBAM device could become a valuable decision support system in various medical fields, including dermatology and plastic surgery, by enabling objective assessment of the mechanical properties of skin. Various diseases could be diagnosed and monitored more effectively, and treatments could be evaluated more accurately. For instance, IBAM could be used, after careful validation, to determine whether the surgical margins taken following tumor resection are sufficient. Furthermore, the technology could also be deployed in the cosmetics industry for assessing product efficacy and providing customer-specific product recommendations.

## 6. Funding

This project has received funding from the European Research Council (ERC) under Horizon 2020 (grant agreement #101147066, project ERC Proof of Concept AssesSkin), from the project OrthAncil (ANR-21-CE19-0035-03), from the project OrthoMat (ANR-21-CE17-0004), from the project IBAM (ANR-24-LCV2-0009) and from the project SkinImpact (ANR-25-CE45-3188). The authors acknowledge the support from CNRS Innovation for the Prematuration project CarImpact. The authors acknowledge the support of Cloé BUFFARD for performing some experimental measurements.

## 7. Authors' Contributions

Conceptualization: A.B., D.PO., MOI.O., P.K., D.PA., CH.FL., R.B., B.H., JP.M. and G.H.; Data Curation: A.B. and D.PO; Formal Analysis: A.B., D.PO. and G.H.; Funding Acquisition: CH.FL., R.B., B.H., JP.M. and G.H.; Investigation: A.B., D.PO., CH.FL., R.B., B.H., JP.M. and G.H.; Methodology: A.B., D.PO., MOI.O, P.K. and G.H.; Project Administration: G.H.; Resources: MOI.O., D.PA., CH.FL., R.B., B.H., JP.M. and G.H.; Software: A.B., P.K. and G.H.; Supervision: G.H.; Validation: A.B., D.PO. and G.H.; Visualization: A.B.; Writing – Original Draft: A.B., D.PO. and G.H.; Writing – Review & Editing: A.B., D.PO., MOI.O., P.K., D.PA., CH.FL., R.B., B.H., JP.M. and G.H.

## 8. Declaration of Conflicting Interests

The authors declare that they have no financial or non-financial interests that are directly or indirectly related to the work submitted for publication.

## 9. Ethical Statement

This animal experimentations followed the French and European Guidelines for care and use of laboratory animals and were approved by the local ethical committee of ComEth CE016 (n°#44667-2023061717481626 v7).